\documentclass[conference,a4paper]{APSIPA2021}
\usepackage{amsmath}
\usepackage{graphicx}
\usepackage{multirow}
\usepackage{threeparttable}
\usepackage{subcaption}
\usepackage{multicol}
\usepackage{array}
\newcolumntype{P}[1]{>{\centering\arraybackslash}p{#1}}

\usepackage{geometry}
\usepackage{fancyhdr}

\fancypagestyle{firststyle}{
  \fancyhf{}
  \fancyhead[C]{2023 Asia Pacific Signal and Information Processing Association Annual Summit and Conference (APSIPA ASC)}
}

\begin{document}

\title{Cross-domain Sound Recognition \\ for Efficient Underwater Data Analysis}

\author{
\authorblockN{
Jeongsoo Park\authorrefmark{1},
Dong-Gyun Han\authorrefmark{2}\authorrefmark{3}\authorrefmark{4}, 
Hyoung Sul La\authorrefmark{3}, 
Sangmin Lee\authorrefmark{1},
Yoonchang Han\authorrefmark{1}, and 
Eun-Jin Yang\authorrefmark{3}
}

\authorblockA{
\authorrefmark{1}
Cochl Inc, USA \\
E-mail: \{jspark, ychan, smlee\}@cochl.ai }

\authorblockA{
\authorrefmark{2}
Division of Ocean Sciences, Korea Polar Research Institute, Incheon, Republic of Korea \\
E-mail: dghandg@gmail.com, \{hsla, ejyang\}@kopri.re.kr
}

\authorblockA{
\authorrefmark{3}
Research Center for Ocean Security Engineering and Technology, Hanyang University ERICA, Republic of Korea \\
}

\authorblockA{
\authorrefmark{4}
Oceansounds Inc, Republic of Korea 
}

}

\maketitle
\thispagestyle{firststyle}
\pagestyle{fancy}

\begin{abstract}
This paper presents a novel deep learning approach for analyzing massive underwater acoustic data by leveraging a model trained on a broad spectrum of non-underwater (aerial) sounds. Recognizing the challenge in labeling vast amounts of underwater data, we propose a two-fold methodology to accelerate this labor-intensive procedure.

The first part of our approach involves PCA and UMAP visualization of the underwater data using the feature vectors of an aerial sound recognition model. This enables us to cluster the data in a two dimensional space and listen to points within these clusters to understand their defining characteristics. This innovative method simplifies the process of selecting candidate labels for further training.

In the second part, we train a neural network model using both the selected underwater data and the non-underwater dataset. We conducted a quantitative analysis to measure the precision, recall, and F1 score of our model for recognizing airgun sounds, a common type of underwater sound. The F1 score achieved by our model exceeded 84.3\%, demonstrating the effectiveness of our approach in analyzing underwater acoustic data.

The methodology presented in this paper holds significant potential to reduce the amount of labor required in underwater data analysis and opens up new possibilities for further research in the field of cross-domain data analysis.
\end{abstract}

\section{Introduction}
Underwater acoustics is a complex and vast field with a wide range of applications, such as ocean survey, marine life monitoring, and submarine detection. In general, the underwater sound data is collected via Passive Acoustic Monitoring (PAM), which is a non-invasive technique aiming to record cetacean sounds. However, the analysis of underwater acoustic data is challenging due to the unique characteristics of underwater sound. The underwater sound propagating through the acoustic waveguide interacts with the boundaries (i.e., reflection and scattering) while also undergoing refraction due to variations in the sound speed in the water medium. Additionally, the kinds of animals that reside underwater and the kinds of artificial noise that humans make are different from the aerial noise. As a result, even if we have a massive sound dataset in advance of the underwater sound analysis, it can be difficult to reuse it directly to analyze the underwater data.

Hence, the studies concerning underwater acoustic data have been focusing on making a sound recognizer from the scratch. Many research efforts used conventional rule-based methods such as Erbe and King's method using information theory \cite{1} and Baumgartner \textit{et al.}'s spectrogram filtering-based method \cite{2}. In Yack \textit{et al.}'s study \cite{3}, six detection algorithms were tested and compared. In other studies including Halkias \textit{et al.}'s \cite{4} and Thomas \textit{et al.}'s work \cite{5}, deep neural network architecture was used. Usman \textit{et al.} summarized the recent research trends and their methods \cite{6}.

To train a neural network model that is capable of recognizing underwater sounds, we need an enormous amount of labeled data. However, labeling underwater acoustic data is a labor-intensive task due to the limited available labeled data and the complexity of the underwater environment. Additionally, the people who can distinguish the underwater sound events are very rare because they need to have expertise in the underwater environments and acoustics.

To overcome this problem, several studies tried to reduce the amount of data required to train a model. Both Shiu \textit{et al.} \cite{7} and Webber \textit{et al.} \cite{8} generalized a model trained on a small underwater data to analyze massive database consists of several years of recording. Thomas \textit{et al.} utilized a transfer learning approach to generalize a neural network model to detect other sources \cite{9}. However, none of the previous studies tried to utilize aerial sounds to analyze the underwater data even though they are easy to collect and guarantee diversity of the data.

In this paper, we propose a novel approach to address the challenges of underwater acoustic data analysis. Our methodology is a cross-domain approach, which is about applying state-of-the-art aerial sound recognition AI to underwater data to accelerate the entire analysis process. Our methodology is two-fold, involving the visualization of underwater data using the feature vectors of a sound recognition model trained on non-underwater sounds and quantitative analysis using this same model.

By using a deep learning model trained on a broad array of aerial sounds, we can navigate around the limited availability of labeled underwater data. Furthermore, visualizing and listening to the clustered data allows for efficient labeling and selection of candidate labels for additional training.

Following this approach, we conducted an in-depth analysis of our model's performance with underwater acoustic data, specifically focusing on the recognition of airgun sounds. The results highlight the model's high precision, recall, and F1 score, demonstrating the effectiveness and robustness of our approach. Through this research, we aim to pave the way for more efficient underwater acoustic data analysis, reducing labor-intensive processes, and fostering further innovations in the field.

The rest of this paper is organized as follows. In Section II, we explain our underwater data and how we collected it. Section III covers our method for the visualization of the underwater data. Section IV provides the empirical measures that confirm the success of our model in detecting underwater sounds. In Section V, we conclude the paper with our final thoughts.

\section{Underwater Dataset and Collection Method}
\subsection{Underwater sound data}
The underwater dataset used in this study was collected in the East Siberian Sea, Arctic Ocean, using a hydrophone over one year \cite{10}. The data collection period spanned from August 2017 to August 2018.

The hydrophone was programmed to record 600 seconds of sound every hour, with a sample rate of 32,768Hz, a common rate for high-quality underwater audio capture. Over the one-year period, this method provided a comprehensive audio snapshot of the underwater environment, capturing a broad range of underwater sound events.

\subsection{Airgun labels}
Among the various sound events in the dataset, specific labels for airgun sounds were included. Airguns, widely used in marine seismic surveys, produce unique acoustic signatures and are among the more recognizable sound events in underwater acoustic data. Unlike other sounds, such as mammal sounds, airgun sounds are steady in terms of the sound pattern. Hence, collecting labels for this sound is relatively straightforward with the help of automated algorithms.

In particular, we focused on the September 2017 data, since most of the airgun sounds occur in the summer season \cite{11}. For the initial identification of airgun sounds in our dataset, we employed a matched filter algorithm as described in \cite{10} and \cite{12}. Following this semi-automated identification process, an expert in underwater acoustics reviewed the potential airgun events. This step ensured the accuracy of the labels, verifying that the sounds identified by the algorithm were indeed airgun sounds. Additionally, the expert examined if there were any missing airgun sounds.

As a result, the dataset we obtained covered a wide range of underwater sounds and also contained a verified set of labeled airgun sounds, serving as a valuable resource for our deep learning model training and evaluation.

\section{Underwater Sound Visualization for Initial Identification}
The manual labor involved in the categorization and identification of large, unlabeled underwater acoustic datasets can be overwhelming. This challenge calls for a method that can aid in the initial identification and selection of underwater sound characteristics without relying on extensive manual labeling. This section introduces our innovative approach to visualizing underwater sounds, which streamlines this process and provides a launching point for further analysis.

Our methodology relies on a deep learning model trained on a broad array of aerial sounds. The basis for this approach is the understanding that a variety of aerial sounds can capture a wide range of acoustic features, which can be extended to the underwater environment.

\subsection{Principal component analysis}
Our process begins by feeding the underwater acoustic data through our trained sound recognition model. The audio data is segmented into one-second snippets, with no overlap between adjacent snippets, to match our model's input requirement of a 22,050Hz audio snippet lasting one second. These segmented chunks are then fed into the model. For this experiment, we removed the final layer, which functions as the classifier, to obtain the feature vectors of 1,280 dimensions. These contain intricate information about the acoustic properties of the underwater sounds. These feature vectors serve as the coordinates in a high-dimensional space where similar sounds cluster together.

However, visualizing and interpreting high-dimensional data is neither intuitive nor straightforward. Therefore, we employ Principal Component Analysis (PCA), which is a dimensionality reduction technique, to project the high-dimensional feature vectors into a 2D space. In the visualized space, clusters of similar sounds emerge based on their acoustic features, making them discernible to the human eye and ear.

This visualization allows researchers to isolate and listen to individual points or clusters within the projected data. By doing so, we can qualitatively assess the sound characteristics of these clusters and choose suitable candidate labels for the sounds within them.

\begin{figure*}[htb]
    \centering
    \begin{subfigure}[b]{0.33\textwidth}
        \includegraphics[width=\textwidth]{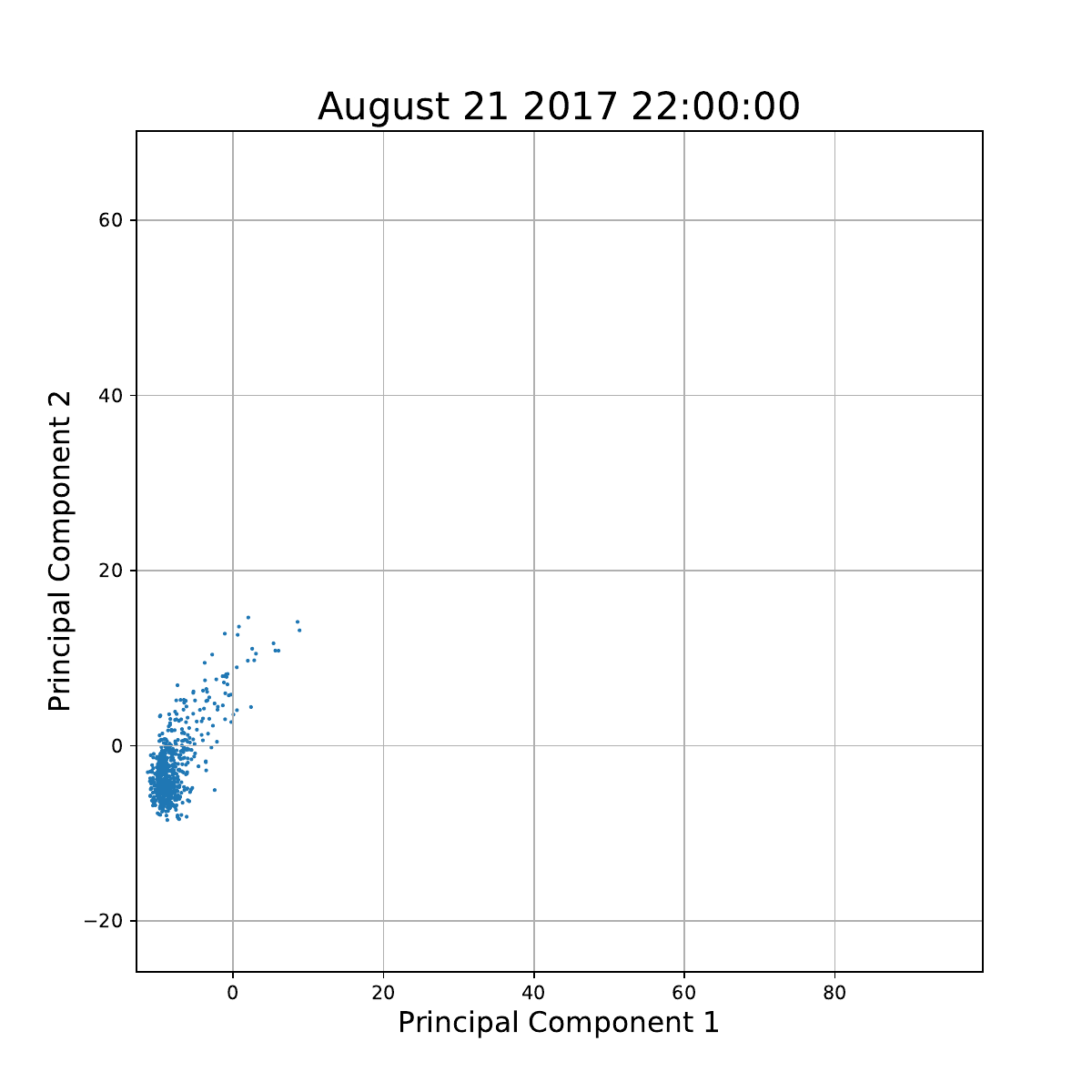}
        \caption{}
        \label{fig:1a}
    \end{subfigure}%
    \begin{subfigure}[b]{0.33\textwidth}
        \includegraphics[width=\textwidth]{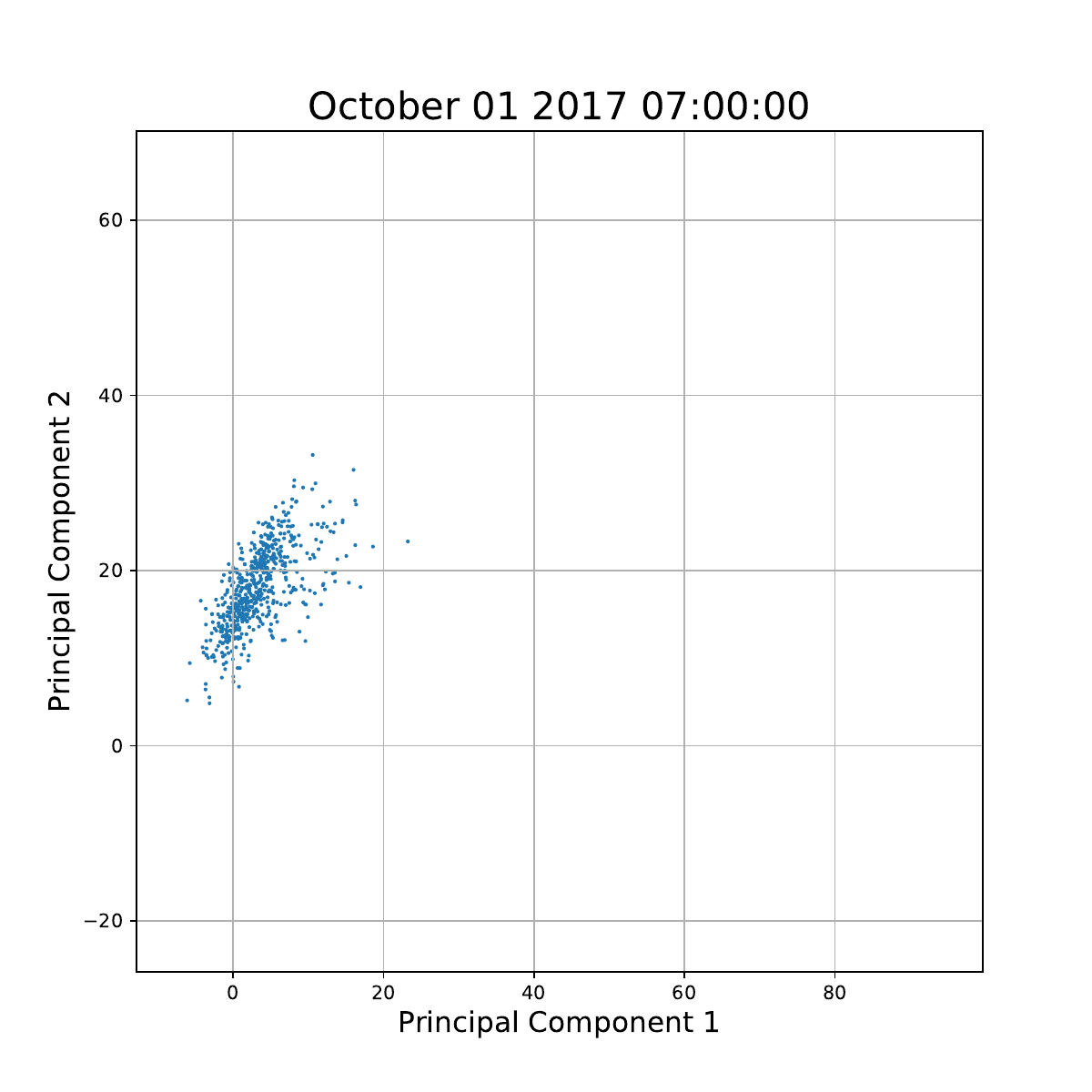}
        \caption{}
        \label{fig:1b}
    \end{subfigure}%
    \begin{subfigure}[b]{0.33\textwidth}
        \includegraphics[width=\textwidth]{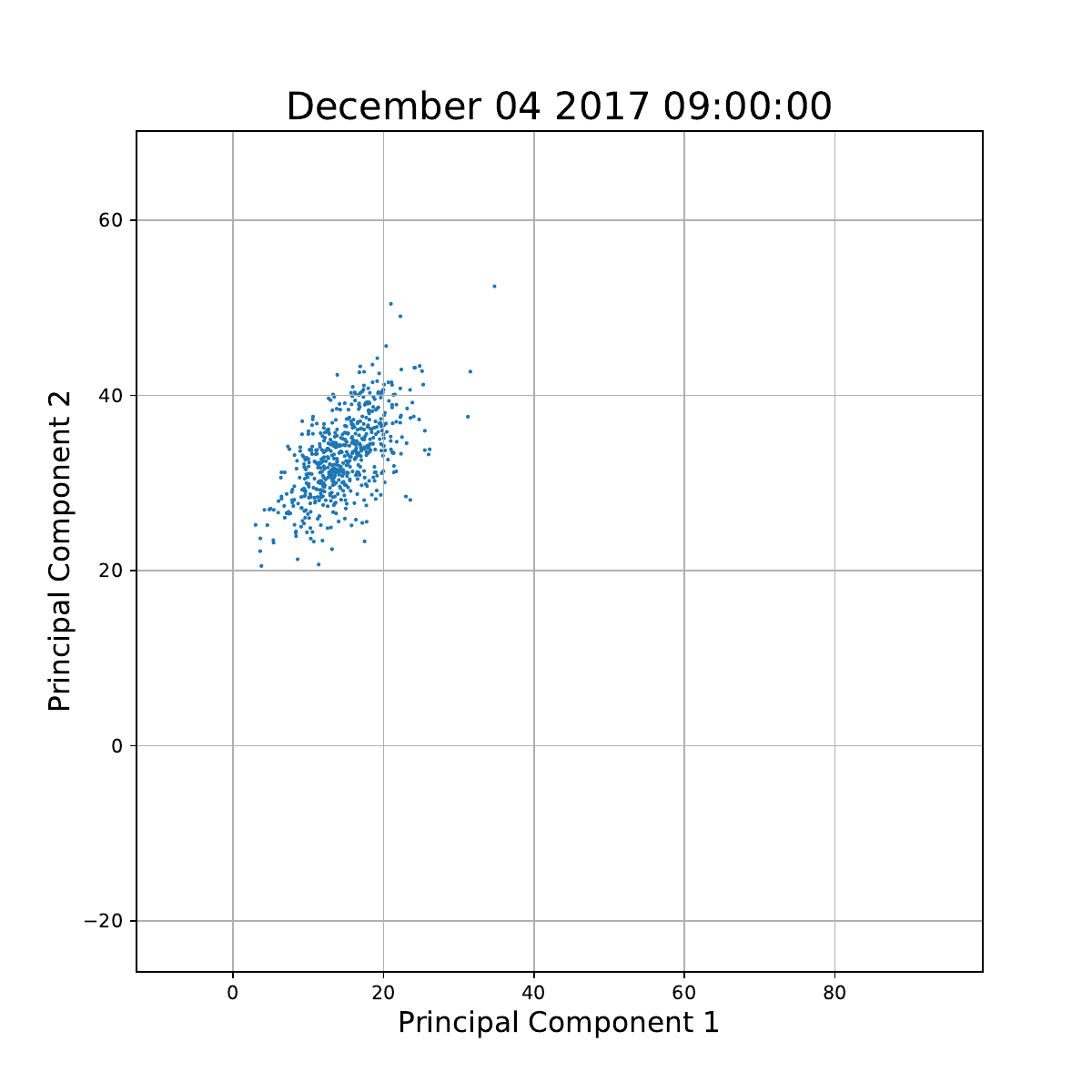}
        \caption{}
        \label{fig:1c}
    \end{subfigure}

    \begin{subfigure}[b]{0.33\textwidth}
        \includegraphics[width=\textwidth]{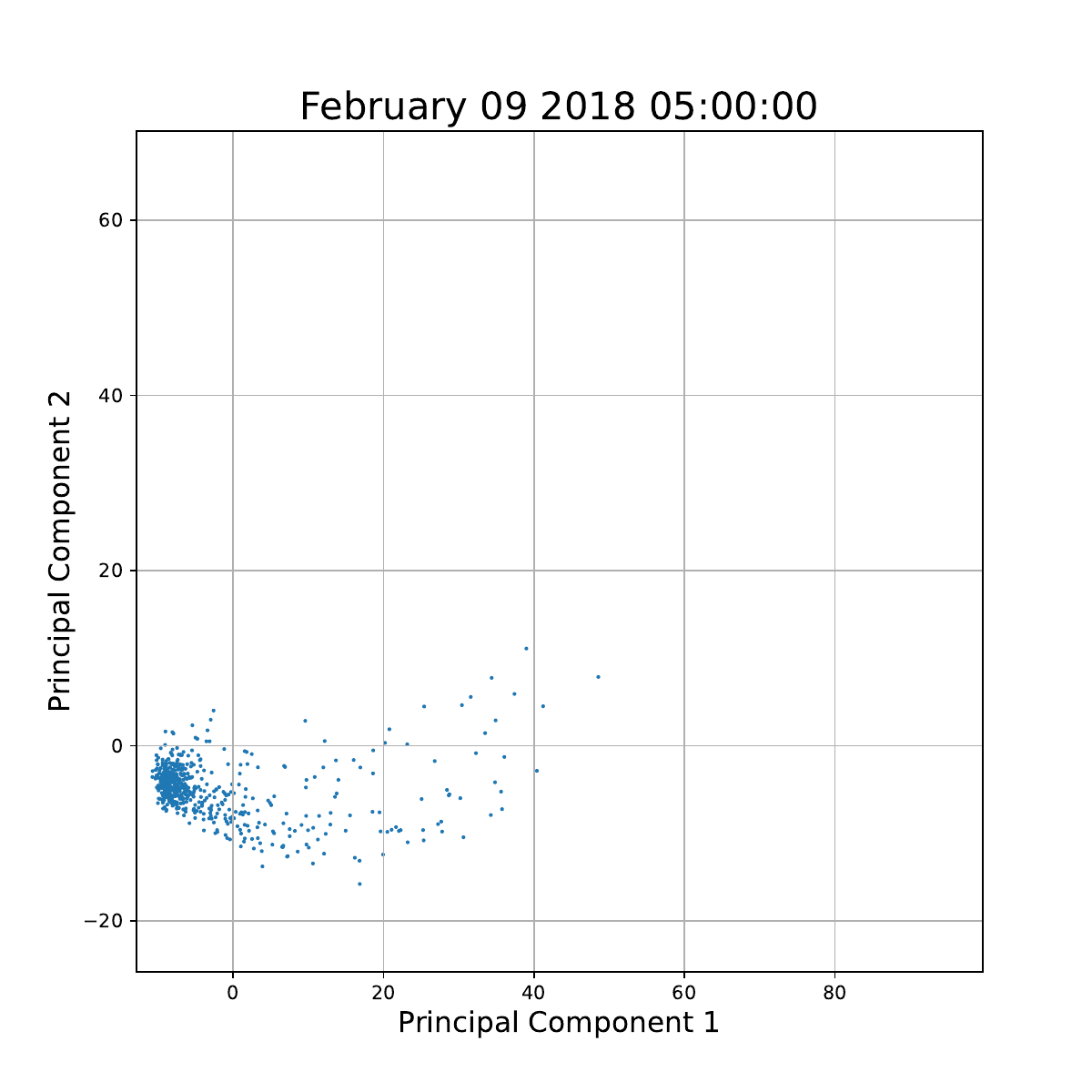}
        \caption{}
        \label{fig:1d}
    \end{subfigure}%
    \begin{subfigure}[b]{0.33\textwidth}
        \includegraphics[width=\textwidth]{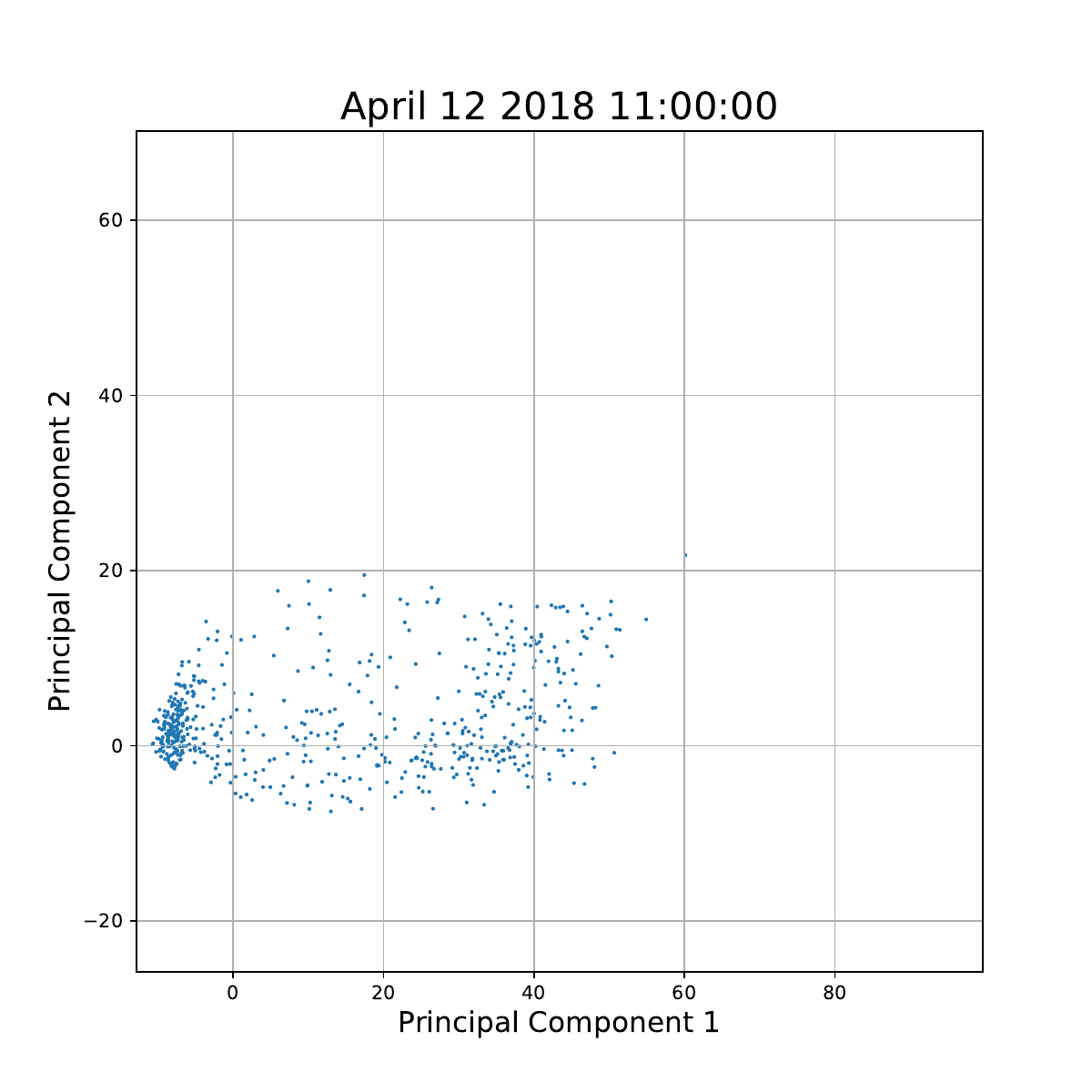}
        \caption{}
        \label{fig:1e}
    \end{subfigure}%
    \begin{subfigure}[b]{0.33\textwidth}
        \includegraphics[width=\textwidth]{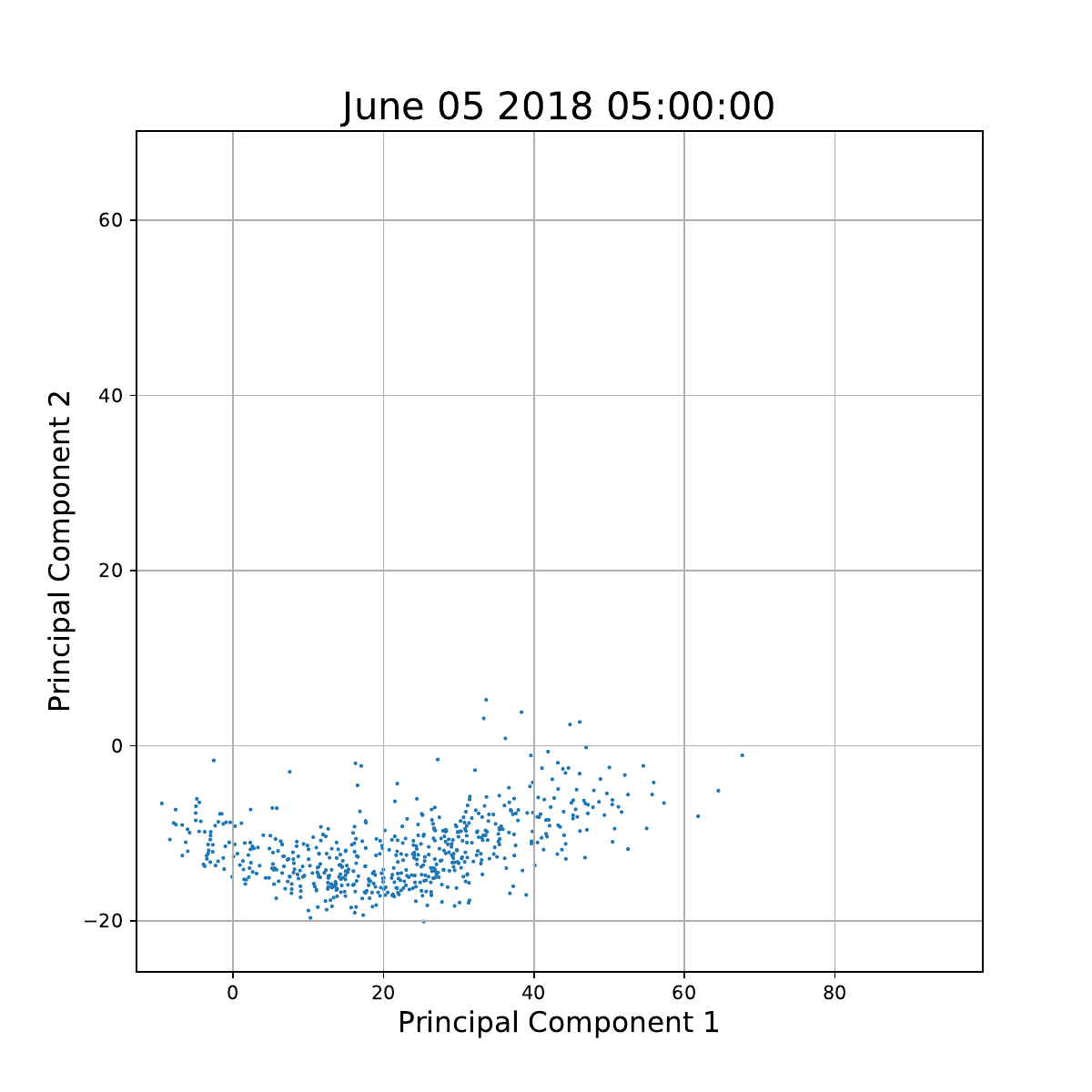}
        \caption{}
        \label{fig:1f}
    \end{subfigure}

    \caption{Visualizing and analyzing the distributions of audio data based on seasonal sound characteristic variations.}
    \label{fig:1}
\end{figure*}

Fig. \ref{fig:1} shows various visualizations of the underwater sounds. Each subfigure represents the data of different dates and times. It can be seen that the data points spread in different locations on the plane. For example, the data points of \textbf{August 21, 2017, 22:00:00} are condensed around (0, 0), as in Fig. \ref{fig:1a}. Meanwhile, those of \textbf{December 04, 2017, 09:00:00} in Fig. \ref{fig:1c} have a higher $Principal Component 2$. In the \textbf{June 05, 2018, 05:00:00} data, we can observe that the data points are spread across various $Principal Component 1$ as in Fig. \ref{fig:1f}.

This difference in distribution may be caused by the sound characteristics, which in turn are influenced by the environmental changes due to the changing seasons. Deeper investigation by an expert revealed that $Principal Component 1$ is related to the presence of animal sounds such as those of bearded seals. Besides, $Principal Component 2$ is associated with sea ice concentration and fluid velocity.

For further analysis, we extracted 84,499 one-second long sound data that had a $Principal Component 1$ larger than 40. Then, the underwater data expert screened out the ones that contain sound events from those which do not. As a result, we obtained the labeled dataset shown in Table \ref{tab:sound_data}. It is interesting to observe that most of the sound events are either bearded seals or airguns. Also, each sound is concentrated in a certain season. For example, all the airgun sounds were detected in September 2017, and bearded seal sounds were detected in either March or May 2018.

\subsection{Uniform Manifold Approximation and Projection}
Uniform Manifold Approximation and Projection, often referred to as UMAP, is a technique to reduce data dimension \cite{13}. It is often used for high-dimensional data visualization. Even though PCA is a simple approach to visualize data, we can further improve the visualization by applying UMAP and obtain further insights.

\begin{figure*}[htb]
    \centering
    \begin{subfigure}[b]{0.5\textwidth}
        \includegraphics[width=\textwidth]{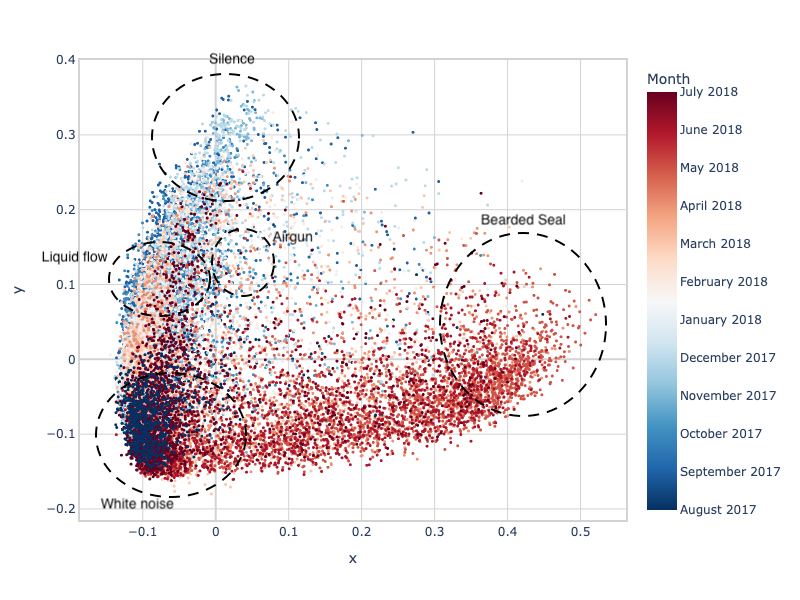}
        \caption{PCA}
        \label{fig:2a}
    \end{subfigure}%
    \begin{subfigure}[b]{0.5\textwidth}
        \includegraphics[width=\textwidth]{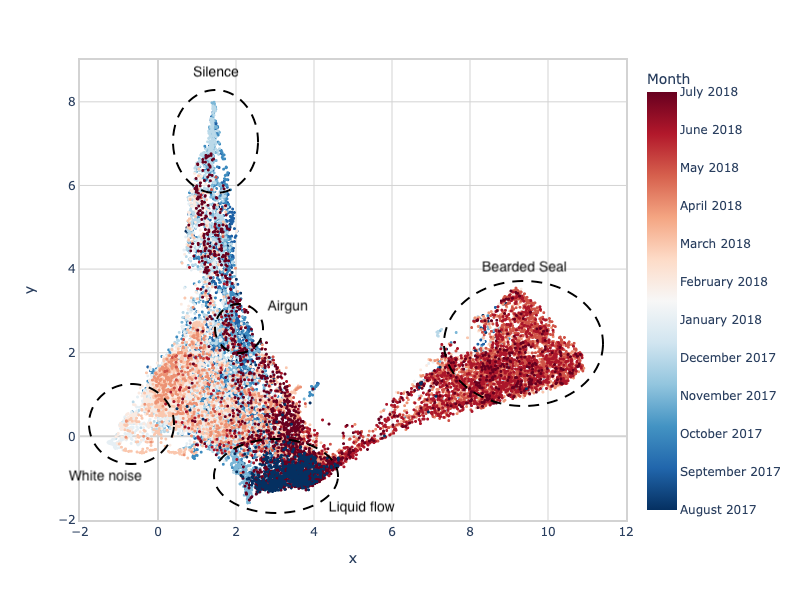}
        \caption{UMAP (n\_neighbors=10) }
        \label{fig:2b}
    \end{subfigure}%

    \caption{Comparison of PCA and UMAP visualization.}
    \label{fig:2}
\end{figure*}

Fig. \ref{fig:2a} and Fig. \ref{fig:2b} show the visualizations using PCA and UMAP, respectively. To reduce computational complexity and enhance graph visibility, we used 25,650 randomly chosen data points, which constitute 0.5\% of the entire data in quantity. It is interesting to see that the data points of a season are gathered together in both plots. In the PCA visualization, the sound cluster boundaries are relatively unclear and the data points are scattered. Based on the clear clusters of data points that represent specific types of sounds, we can argue that the UMAP algorithm better visualizes our data.

The sound visualization method drastically simplifies the process of labeling, offering an intuitive, visual, and auditory understanding of the characteristics and distribution of the underwater acoustic data. In the subsequent section, we will outline the process of training our new deep learning model with selected candidate labels and the aerial dataset, followed by a rigorous quantitative analysis of the model's performance.

\begin{table}
\centering
\caption{Sound Types and Characteristics of the data with $Principal Component 1 > 40$.}
\begin{tabular}{|P{1.8cm}|c|P{3.5cm}|}
\hline
Sound Type  & Number of Data    & Occurrence Date \\
\hline
Bearded Seal & 1033 & Mar. or May 2018 \\
\hline
Walrus       & 9    & Sept. or Oct. 2017 \\
\hline
Airgun       & 275  & Sept. 2017 \\
\hline
Sea ice      & 1    & Aug. 25 2017 00:00:00 \\
\hline
Whales       & 12   & Sept. 11 or 19 2017 \\
\hline
Mammal       & 7    & Sept. 23 2017 17:00:00 \\
\hline
\end{tabular}
\label{tab:sound_data}
\end{table}

\section{Quantitative Analysis of Underwater Sound Recognition}
\subsection{Model training}
This section applies quantitative analysis to underwater data, focusing specifically on the recognition of airgun sounds. In the previous chapter, candidate labels for the underwater sounds were selected based on the results of our visualization and on listening to the clustered data. Bearded seal and airgun labels served as starting points for a more comprehensive training process. They were combined with the aerial sound dataset with the aim of creating a robust model capable of recognizing a variety of underwater sounds. Other underwater sound event labels, such as whales, were discarded due to the small quantity of data available for training.

The trained model employs a Convolutional Neural Network (CNN)-based architecture with an argmax classifier. An audio snippet with a sample rate of 22,050Hz and 22,050 samples in length is fed into the model as input and transformed into a melspectrogram. Because the original data has a sample rate of 32,768Hz, downsampling to 22,050Hz is necessary prior to training.

A significant advantage of our approach is its ability to work with relatively small amounts of labeled underwater data due to the use of the aerial sound dataset. This substantially reduces the burden of manual labeling and allows us to quickly move into the training phase. Cochl's aerial sound dataset, which consists of 586,343 labels across 242 classes, helps the new model learn the feature representation of the sounds. Both aerial and underwater sound data are split into train, validation, and test sets, with the validation and test sets each containing 10\% of the data.

\subsection{Internal evaluation}
After training, a comprehensive quantitative analysis of the model's performance is conducted, focusing specifically on the recognition of airgun sounds. Precision, recall, and the F1 score are used as the primary metrics for evaluating the performance of our model.

Precision measures the model's ability to correctly identify positive instances out of all instances it classified as positive. Recall, on the other hand, determines the model's ability to identify all positive instances from the total actual positives. The F1 score harmonizes these two metrics, providing an overall assessment of the model's accuracy in recognizing airgun sounds.

In the first part of the evaluation, an internal evaluation is performed using the test split of the dataset. It consists of 28 labels of airgun, 103 labels of bearded seal, and 173,844 labels of the rest of the aerial sound data. Notably, the underwater ambient noise data is not included in the test set.

Interestingly, both the recall and precision of airgun and bearded seal sounds are estimated as 1.0, which in turn causes the F1 scores also to be 1.0. This high performance can be attributed to the fact that the training and test data share the same characteristics because they are randomly split. Also, the exclusion of ambient underwater sound from the dataset may have made the evaluation easier because it is closer to the target sounds than the aerial sounds.

\subsection{External evaluation}
The second part of the evaluation uses the airgun labels mentioned in Section II-B. Along with the labeled airgun sounds, the remaining underwater sounds are also used to check for false positives. Preliminarily, labels overlapping with the training data are eliminated to avoid the influence of the training and internal test data on the performance. As a result, the total count of the airgun data is 20,397, whereas that of the ambient underwater data is 259,238. The sound snippets are cut into 1-second chunks without overlap with the neighboring chunk and fed into the model.

We initially used the argmax criteria, as in the training phase. However, we also observed changes in performance according to the threshold level. As the model returns the probability for three classes, the probability of the detected class has to exceed 0.3333. Hence, lowering the threshold with the threshold-based detection criteria increases the recall.

\begin{figure*}[htb]
    \centering
    \includegraphics[width=\textwidth]{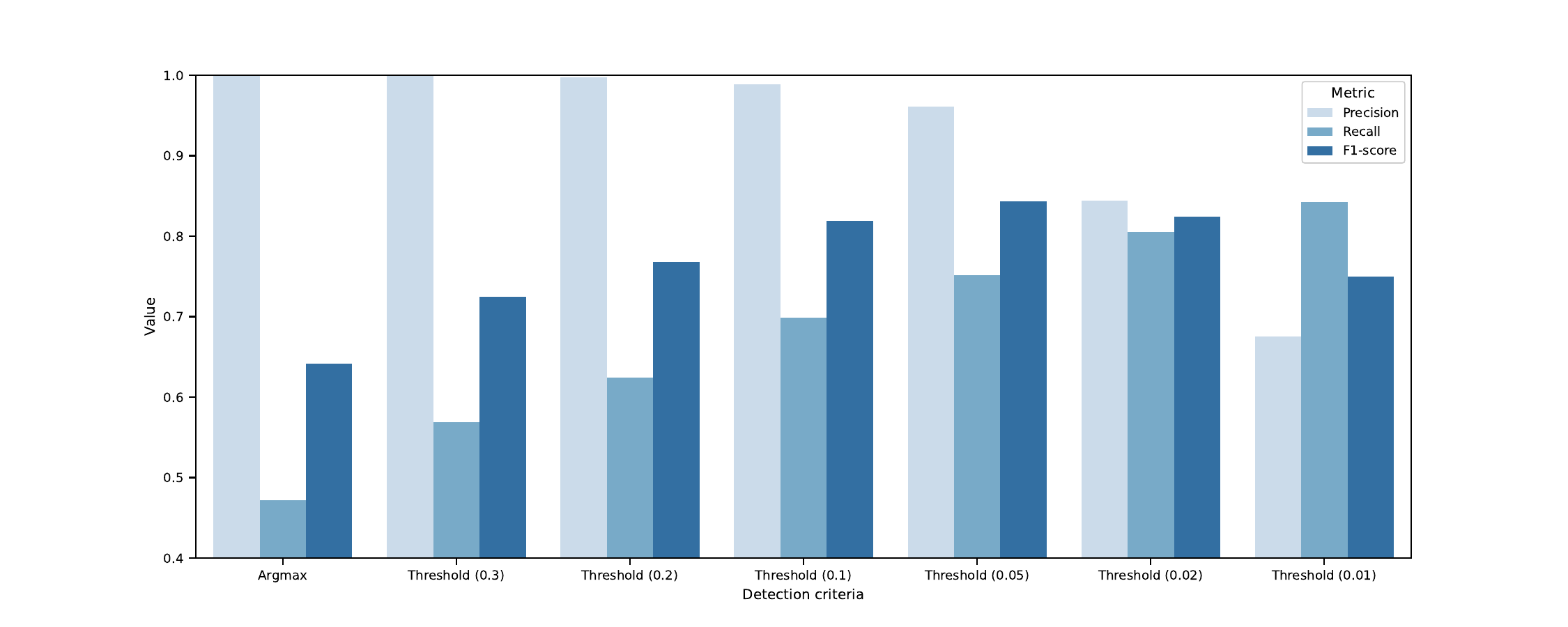}
    \caption{Performance metrics for airgun detection: Precision, Recall, and F1-score. Optimal F1-score is achieved when the detection threshold is established at 0.05.}
    \label{fig:3}
\end{figure*}

Fig. \ref{fig:3} shows the evaluation results. The precision under the argmax criteria is noticeably high, with the model only detecting 4 false positives out of a total of 259,238 snippets. These misidentified chunks were walrus moan sounds, which may be hard to distinguish from airgun sounds for non-experts. Nevertheless, the model struggled to achieve a high recall rate, potentially attributable to the constrained volume and diversity of the data. The lack of underwater ambient sound data in the training set might have also contributed to this issue. Moreover, the model's sensitivity to the start time of input audio chunks could have complicated results, as airgun sounds are short and can be challenging to identify if split between two consecutive chunks. Using the threshold-based detection criterion, the recall rate escalates as the threshold descends, thereby accommodating more false positives. The optimal F1-score can be achieved at a threshold value of 0.05.

Upon observing the threshold and probability values, we can infer that the large amount of aerial data might have influenced the probability of the airgun sounds. This is particularly relevant considering the argmax layer's characteristic that normalizes the sum of probabilities to 1.0, which could have magnified the effect.

In conclusion, our deep learning model showcases an effective and efficient method for recognizing underwater sounds. This quantitative analysis underscores the value of our approach in streamlining labor-intensive processes in underwater acoustic data analysis, and it emphasizes its potential for future improvements. However, some types of post-processing, such as threshold optimization, may be necessary to maintain a high recall rate.

\section{Conclusion}
In this paper, we have presented a novel approach to accelerate underwater sound analysis by leveraging an aerial sound recognition model. Successful visualization and labeling of underwater sound data were achieved, and a sound recognition model that achieves F1 scores of 1.0 in internal evaluation was trained. In external evaluation, although the overall probability of airgun detection was low, we managed to obtain an F1 score of 84.3\% when we set the threshold to a lower value. This study is the first to utilize an aerial sound recognition model to accelerate the process of underwater sound analysis. Our approach holds significant potential to improve the efficiency of underwater sound analysis substantially. This efficiency could be utilized to a variety of applications, such as monitoring marine life and detecting underwater threats.



\begin{thebibliography}{1}

\bibitem{1}
C. Erbe, and A. R. King, ``Automatic detection of marine mammals using information entropy,'' The Journal of the Acoustical Society of America, Vol. 124, Issue 5, pp. 2833-2840, 2008.

\bibitem{2}
M. F. Baumgartner, and S. E. Mussoline, ``A generalized baleen whale call detection and classification system,'' The Journal of the Acoustical Society of America, Vol. 129, Issue 5, pp. 2889-2902, 2011.

\bibitem{3}
T. M. Yack, J. Barlow, M. A. Roch, H. Klinck, S. Martin, D. K. Mellinger, and D. Gillespie, ``Comparison of beaked whale detection algorithms,'' Applied Acoustics, Vol. 71, Issue 11, pp. 1043-1049, 2010.

\bibitem{4}
X. C. Halkias, S. Paris, and H. Glotin, ``Classification of mysticete sounds using machine learning techniques,'' The Journal of the Acoustical Society of America, Vol. 134, Issue 5, pp. 3496-3505, 2013.

\bibitem{5}
M. Thomas, B. Martin, K. Kowarski, B. Gaudet, and S. Matwin, ``Marine mammal species classification using convolutional neural networks and a novel acoustic representation,'' In Machine Learning and Knowledge Discovery in Databases: European Conference, ECML PKDD 2019, Würzburg, Germany, September 16–20, 2019, Proceedings, Part III, pp. 290-305, Springer International Publishing, 2020.

\bibitem{6}
A. M. Usman, O. O. Ogundile, and D. J. Versfeld, ``Review of automatic detection and classification techniques for cetacean vocalization,'' IEEE Access, Vol. 8, pp. 105181-105206, 2020.

\bibitem{7}
Y. Shiu, K. J. Palmer, M. A. Roch, E. Fleishman, X. Liu, E. M. Nosal, and H. Klinck, ``Deep neural networks for automated detection of marine mammal species,'' Scientific Reports, Vol. 10, Issue 1, Article 607, 2020.

\bibitem{8}
T. Webber, D. Gillespie, T. Lewis, J. Gordon, T. Ruchirabha, and K. F. Thompson, ``Streamlining analysis methods for large acoustic surveys using automatic detectors with operator validation,'' Methods in Ecology and Evolution, Vol. 13, Issue 8, pp. 1765-1777, 2022.

\bibitem{9}
M. Thomas, B. Martin, K. Kowarski, B. Gaudet, and S. Matwin, ``Marine mammal species classification using convolutional neural networks and a novel acoustic representation,'' In Machine Learning and Knowledge Discovery in Databases: European Conference, ECML PKDD 2019, Würzburg, Germany, September 16–20, 2019, Proceedings, Part III, pp. 290-305, Springer International Publishing, 2020.

\bibitem{10}
D. G. HAN, et al. ``Effects of geophony and anthrophony on the underwater acoustic environment in the East Siberian Sea, Arctic Ocean,'' Geophysical Research Letters, 2021, 48.12: e2021GL093097.

\bibitem{11}
S. B. Blackwell , C. S. Nations, T. L. McDonald, A. M. Thode, D. Mathias, K. H. Kim, C. R. Greene Jr, A. M. Macrander, ``Effects of airgun sounds on bowhead whale calling rates: Evidence for two behavioral thresholds,'' \emph{PloS one,} vol. 10, no. 6, 2015.

\bibitem{12}
C. W. Clark, P. Marler, and K. Beeman, ``Quantitative analysis of animal vocal phonology: an application to swamp sparrow song,'' Ethology, Vol. 76, Issue 2, pp. 101-115, 1987.


\bibitem{13}
L. McInnes, J. Healy, and J. Melville, ``Umap: Uniform manifold approximation and projection for dimension reduction,'' arXiv preprint arXiv:1802.03426 (2018).

\bibitem{14}
C. W. Clark, P. Marler, and K. Beeman, ``Quantitative analysis of animal vocal phonology: an application to swamp sparrow song,'' Ethology, Vol. 76, Issue 2, pp. 101-115, 1987.
\bibitem{15}
P. J. Dugan, A. N. Rice, I. R. Urazghildiiev, and C. W. Clark, ``North Atlantic right whale acoustic signal processing: Part I. Comparison of machine learning recognition algorithms,'' In 2010 IEEE Long Island Systems, Applications and Technology Conference, pp. 1-6, IEEE, 2010.
\bibitem{16}
D. Gillespie, M. Caillat, J. Gordon, and P. White, ``Automatic detection and classification of odontocete whistles,'' The Journal of the Acoustical Society of America, Vol. 134, Issue 3, pp. 2427-2437, 2013.
\bibitem{17}
D. K. Mellinger, S. W. Martin, R. P. Morrissey, L. Thomas, and J. J. Yosco, ``A method for detecting whistles, moans, and other frequency contour sounds,'' The Journal of the Acoustical Society of America, Vol. 129, Issue 6, pp. 4055–4061, 2011.

\bibitem{18}
X. Lurton. ``An introduction to underwater acoustics: principles and applications,'' London: springer, 2002.

\end{thebibliography}
\end{document}